\documentstyle[12pt,preprint,aps,tighten,epsfig,floats]{revtex}
\input{epsfig}
\begin{document}
\def\Tr{\,{\rm Tr}\,}
\def\beq{\begin{equation}}
\def\eeq{\end{equation}}
\def\beqa{\begin{eqnarray}}
\def\eeqa{\end{eqnarray}}
\begin{titlepage}
\vspace*{-1cm}
\noindent
\phantom{bla}
%\hfill{$\scriptstyle{\rm UMHEP-453}$}
\\
\vskip 2.0cm
\begin{center}
{\Large {\bf $K \to \pi\pi$  Phenomenology in 
the Presence of Electromagnetism}}
\end{center}
\vskip 1.5cm
\begin{center}
{\large Vincenzo Cirigliano, John F. Donoghue 
and Eugene Golowich} \\
\vskip .15cm
%$^a$ Dipartimento di Fisica dell'Universit\`a and I.N.F.N. \\
%Via Buonarroti,2 56100 Pisa (Italy) \\
%\vskip .15cm
Department of Physics and Astronomy \\
University of Massachusetts \\
Amherst MA 01003 USA\\
vincenzo@het2.physics.umass.edu \\
donoghue@physics.umass.edu \\
gene@physics.umass.edu \\

\vskip .3cm
\end{center}
\vskip 1.5cm
\begin{abstract}
%\begin{center}
We describe the influence of electromagnetism on the phenomenology
of $K \to \pi\pi $ decays. This is required because the 
present data were analyzed without inclusion of electromagnetic
radiative corrections, and hence contain several ambiguities
and uncertainties which we describe in detail. Our presentation
includes a full description of the
infrared effects needed for a new experimental analysis. It also
describes the general treatment of final state interaction phases, 
needed because Watson's theorem is no longer valid in the 
presence of electromagnetism. The phase of the isospin-two 
amplitude ${\cal A}_2$ may be modified by $50 \to 100~\%$.  
We provide a tentative analysis 
using present data in order to illustrate the sensitivity to 
electromagnetic effects, and also discuss how the standard treatment of
$\epsilon' /\epsilon$ is modified.
%\end{center}
\noindent 
%Electromagnetism changes the description of 
%rescattering phases in $K \to \pi\pi$ transitions, as 
%Watson's theorem is no longer valid.  We describe these 
%new features.  We show that the electromagnetic phase 
%change in the ${\cal A}_2$ amplitude is enhanced by a factor 
%of ${\cal A}_0/{\cal A}_2 \simeq 22$, and that this changes 
%the phase of ${\cal A}_2$ by $50 \to 100~\%$.  This helps 
%ameliorate a longstanding puzzle in the data.  We also reanalyze 
%CP-violation phenomenology by including this effect, which 
%produces a shift in the phase of $\epsilon'$ comparable to 
%the present uncertainty.
\end{abstract}
\vfill
\end{titlepage}

\section{introduction}

In this paper we address the effect of electromagnetism on the 
phenomenology of $K \rightarrow \pi \pi$ decays. 
Our previous work of Refs.~\cite{em1,em2,em3} has
dealt mainly with the determination of structure dependent EM
effects on the $K \rightarrow \pi \pi$ amplitudes. It is the aim of
the present work to attempt a complete phenomenological analysis. 
We start by briefly reviewing the
standard treatment of $K \rightarrow \pi \pi$ decay amplitudes,
pointing out that there is room for potentially important isospin
breaking effects.  We then focus on the effect of electromagnetic 
interactions (EM); we enumerate the main new features due to EM and
describe their impact on the parameterization of the $K \rightarrow 
\pi \pi$ decay amplitudes.  Our quantitative analysis begins in 
Section~\ref{IRB} where we take up the problem of the infrared 
divergences which are induced by electromagnetism. We provide a 
complete description suitable for use in an experimental analysis 
in Section~\ref{full}.
We also point out that the data on $K \rightarrow \pi \pi$ do not lead
to a direct extraction of the strong phase shift difference $\delta_0
- \delta_2$, because electromagnetism changes the rescattering phases
of the amplitudes. Already a perturbative calculation has provided
clues for the presence of sizeable effects~\cite{em2}.  To account for
this effect more generally, we provide in Section~\ref{phases} a
suitable extension of Watson's theorem, obtained after writing and
solving the unitarity constraints in presence of isospin-breaking
interactions. Our goal throughout the paper is to relate the theoretical
and experimental issues of these decays, in the hope that a future 
experimental analysis will be undertaken to resolve the substantial
experimental uncertainties.

The reason why present experimental information is not 
adequate is that most of the data was analyzed without
the inclusion of electromagnetic radiative corrections. This means
that the data in the Particle Data Tables is not fully reliable. Moreover,
for certain quantities (namely $\delta_0 - \delta_2$ and the $I = 2$ 
amplitude $A_2$) this uncertainty is enhanced
by a $\Delta I=1/2$ enhancement factor of $22$.  
In Section~\ref{fit} we demonstrate these effects by giving
an illustrative data analysis, trying our best to interpret the
present data.
This step is necessarily tentative and likely partially incorrect, as it
requires knowledge of the experimental procedure used to deal with
soft photons when measuring the branching ratios. In the absence of 
detailed information from the Particle Data Group (PDG), we adopt the 
simplest theoretical framework, not necessarily corresponding to the 
real experimental setup. Within this simple framework we are 
able to show that the extraction of EM free quantities is quite sensitive 
to the treatment of infrared photons.
% $\delta_0 - \delta_2$  on the parameter related to the experimental cuts. 

The other interesting phenomenological issue concerns the effect of
electromagnetism on CP-violating observables. In Section~\ref{CPV} we
discuss the impact of our findings on theoretical
predictions for $\epsilon ' / \epsilon $. In particular, we 
provide a quantitative estimate for the parameter $\Omega^{\rm EM}$, the
effect of a $\Delta I = 5/2$ interaction, and the impact of the new
rescattering effects on the phase of $\epsilon '$.  We conclude the 
paper with a summary of our findings in Section~VII.

\subsection{Standard Phenomenology for $K \rightarrow \pi \pi$ Decays}

Let us start from the conventional phenomenological analysis of the
 decay amplitudes.  There are three physical $K \to
\pi\pi$ decay amplitudes,\footnote{The invariant amplitude ${\cal A}$
is defined via $_{\rm out}\langle \pi\pi|K\rangle_{\rm in} = i
(2\pi)^4 \delta^{(4)}(p_{\rm out} - p_{\rm in}) \left( i {\cal
A}\right)$.}
\beq
{\cal A}_{K^0 \to \pi^+ \pi^-} \equiv {\cal A}_{+-}\ ,
\qquad
{\cal A}_{K^0 \to \pi^0 \pi^0} \equiv {\cal A}_{00}\ , \qquad
{\cal A}_{K^+ \to \pi^+ \pi^0} \equiv {\cal A}_{+0}\ \ .
\label{cap0e1}
\eeq
We consider first these amplitudes in the limit of exact isospin
symmetry and then identify which modifications must occur in the
presence of electromagnetism. 
In the $I = 0,2$ two-pion isospin basis, the physical amplitudes are
parameterized as:
\beqa
{\cal A}_{+-} \ &=&\  A_0 \, e^{i \delta_0} + \sqrt{1\over 2} A_2 \, e^{i
\delta_2} \ \ ,
\nonumber \\
{\cal A}_{00} \ &=& \ A_0 \, e^{i \delta_0} - \sqrt{2} A_2 \,
e^{i \delta_2} \ \ ,
\label{cap0e2} \\
{\cal A}_{+0} \ &=& \ {3 \over 2} A_2 \, e^{i \delta_2} \ \ .
\nonumber
\eeqa
In the above, $A_{0,2}$ are the $\Delta I = 1/2, 3/2$ transition 
amplitudes corresponding to the $\pi \pi$ final states with isospin 
equal to $0,2$ respectively. They are real in the limit of CP
conservation. The $\delta_{I}$ are the $I = 0,2$ $\pi \pi$ scattering
S-wave phase shifts at center of mass energy equal to the kaon
mass. They enter the parameterization as prescribed by unitarity (the
Fermi-Watson theorem).  Knowledge of the experimental branching
ratios~\cite{PDG} allows one to use Eqs.~(\ref{cap0e2}) to extract the
isospin amplitudes.  Neglecting the small $CP$-violation effect, we
find\footnote{Knowledge of the phase difference $\delta_0 - \delta_2$
from other determinations poses the constraint $\cos (\delta_0 -
\delta_2) > 0$, implying that $A_0 A_2 > 0$.  In this paper, we take
$A_0$ and $A_2$ as positive numbers.}
\begin{eqnarray}
 A_{0} & = & (5.458   \pm   0.012) \times 10^{-7} M_{K^{0}} \ , \nonumber \\
 A_{2} & = & (0.2454  \pm   0.0010) \times 10^{-7} M_{K^{0}} \ , 
\label{cap0e3}  \\
 \delta_{0} - \delta_{2} & = & (56.7  \pm  3.8)^{o} \ . \nonumber
\end{eqnarray}
Throughout we express the $K \rightarrow \pi \pi$ amplitudes 
in units of $10^{-7} M_{K^0}$, with $M_{K^0} = 0.497672$ GeV.
 
A careful inspection of these phenomenological results reveals some
inconsistencies with other existing pieces of phenomenology and
theoretical analysis. These clues seem to suggest that isospin
breaking effects (like EM) can play an important role in the 
phenomenology of $K \rightarrow  \pi \pi$ decays. 
The considerations we present below apply to all isospin
breaking interactions.  In this category one includes both EM and {\em
strong} isospin breaking, produced by the difference in the up and
down quark masses. 
In this work we are concerned exclusively with EM effects 
(also analyzed in Refs.~\cite{old,der,Vienna00}).  
For treatments of strong isospin breaking effects  
see Refs.~\cite{isob,isobnew}.

Isospin breaking interactions will in general mix the amplitudes $A_0$
and $A_2$, thus generating potentially big corrections to $A_2$,
proportional to $A_0 \cdot \alpha / \pi$.  A related issue concerns
the presence of a $\Delta I = 5/2$ component in the interaction. This
problem has recently received attention in Ref.~\cite{valencia}.  A
$\Delta I = 5/2$ component will distinguish between the amplitudes $A_2$
entering in the $K^0$ and $K^+$ decays.  The expression of $A_0, A_2,
A_{2}^{+}$ in terms of $A_{\Delta I}$ is given by:
\begin{eqnarray}
 A_{0} & = &  A_{1/2}  \ ,  \nonumber \\
 A_{2} & = &   A_{3/2}  + A_{5/2} \ , \label{cap0e4}  \\
 A_{2}^{+} & = & A_{3/2}  -  2/3 \ A_{5/2} \ . \nonumber  
\end{eqnarray}
We expect the dominant $\Delta I =
5/2$ effect to arise by combining the large $\Delta I = 1/2$ weak
interaction with the $\Delta I = 2$ component of the electromagnetic
interaction. A combination of the $\Delta I = 3/2$ 
hamiltonian with the $\Delta I = 1$ interaction proportional to 
$m_u - m_d$ is also expected to contribute. 
However, its effect is expected to be doubly suppressed 
(by the $\Delta I = 1/2$ rule and the smallness of $m_u - m_d$). 

A further problem with the isospin analysis is revealed by looking at the
extracted phase shifts.  The value $\delta_{0} - \delta_{2} = (56.7
\pm 3.8)^{o}$ obtained from kaon decay data has to be compared with
information coming from other sectors of low energy phenomenology.  In
particular, the value extracted from 
a dispersive treatment of $\pi \pi$ scattering data is 
%$\delta_{0} - \delta_{2} = (46.3 ^{+2.7} _{- 4.0} )^{o} $
$\delta_{0} - \delta_{2} = (45.2 \pm 1.3 ^{+4.5}_{-1.6} )^{o}$~\cite{acgl},  
and the prediction of ChPT~\cite{GassMeiss} is $\delta_{0} -
\delta_{2} = (45 \pm 6)^{o} $. These two determinations are 
mutually compatible. 
However, there is a sizeable discrepancy with the result obtained 
in the fit to $K \rightarrow \pi \pi$.  This can be ascribed to 
isospin breaking and to a non-vanishing $A_{5/2}$. 

The above mentioned issues also affect a proper theoretical 
understanding of the direct CP-violation parameter $\epsilon
'/\epsilon$.  In particular, the leakage of the octet amplitude in
$A_{2}$ (due to isospin breaking effects) brings an extra
contribution to the CP-violating phase of $A_{2}$. In the literature
only the leakage due to $m_{u} - m_{d}$ isospin breaking effects has
been analyzed, and is found to be numerically 
important \cite{isob,isobnew} .
Moreover, the presence of a $\Delta I = 5/2$
amplitude introduces an extra term in the usual formulae for 
$\epsilon '$.  Finally, understanding the issue regarding the phase shift
$\delta_{0} - \delta_{2}$ will provide a better theoretical
determination of the phase of $\epsilon '$.

The above considerations call for a careful analysis of 
electromagnetic effects on $K \rightarrow \pi \pi$ decays.

\subsection{Electromagnetism and the $K \to \pi\pi$ Amplitudes}

\begin{figure}
\centering
\begin{picture}(200,80)  
\put(50,20){\makebox(100,60){\epsfig{figure=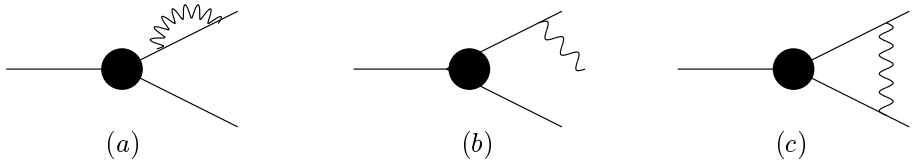,height=0.8in}}}
\end{picture}
\caption{Electromagnetic contributions relevant for the infrared 
behavior.  \label{cap0fig1}}
\end{figure}

One can summarize the effects of electromagnetism on $K \rightarrow 
\pi \pi$ amplitudes as follows: 

\begin{enumerate}
\item First of all one has to deal with universal infrared (IR)
effects, due to photons of long wavelength. These effects are common
to all processes with charged external particles and do not depend on
details concerning the original interaction.  They are represented
diagrammatically in Fig.~\ref{cap0fig1}, where the dark blob is seen
as pointlike by the infrared photons.  This class of contributions
provides a Coulomb final state interaction (FSI) phase and gives
rise to IR divergent amplitudes. Such infrared divergences have to be
canceled by considering the effect of soft real photons
(Fig.~\ref{cap0fig1}(b)).  As we shall show in Sections \ref{IRB} and
\ref{fit}, these effects can be described as modifying 
the phase space factor rather than
producing effects on the amplitudes themselves.
In order to perform an accurate phenomenological study, it is 
important to include this class of radiative corrections in the 
experimental analysis of the branching ratios (See Sect. \ref{fit} for 
details). 

\item There are structure dependent effects, sensitive to the form of the 
original interaction. These are hidden in Fig.~\ref{cap0fig1}  within 
the large dark vertices. 
We consider only corrections induced by the dominant (octet) part 
of the weak hamiltonian.
% (see Fig.\ref{cap3fig2}) 
These produce shifts in the isospin amplitudes, and are responsible 
for possible large contaminations of $A_{2}$.  They also generate 
a $\Delta I = 5/2$ amplitude.  We have calculated these effects in 
ChPT and in dispersive matching.~\cite{em2,em3} 

\item Finally EM affects the final state interaction. 
As a consequence the unitarity relations, determining the
rescattering phases, are altered. The main modifications are due to the
opening of the $\pi \pi + n \gamma$ intermediate channels and the
possibility of mixing between two-pion states in isospin $I = 0$ and
$I = 2$. These new features imply modifications to the unitarity
parameterization, governed by an extension of Watson's theorem 
that we shall discuss at length in Sect.~IV. 
\end{enumerate}

Clearly the items enumerated above are intertwined and affect
in various respects the way one analyzes the $K \rightarrow \pi \pi$
amplitudes. It is important to note that - although being
perturbatively small - the new interaction considered breaks the
original isospin symmetry on which the parameterization of $K
\rightarrow \pi \pi $ amplitudes is based.  Therefore, to perform a
complete analysis of EM effects to $K \to \pi \pi$ decays, one must
understand how to parameterize the above mentioned effects.

\section{Infrared Behavior and Isospin Amplitudes}
\label{IRB}

\subsection{Defining Infrared Finite Amplitudes}

Let us start by summarizing the regularization and removal of infrared
divergences. These arise in perturbation theory through diagrams in
which virtual photons connect external charged legs.  The classical
works of Refs.~\cite{{F61},{Weinberg65}} show how to sum the infrared
singularities to all orders in perturbation theory and isolate an
infrared finite amplitude. We begin by reviewing the content of these
works. Let us introduce an IR regulator $\lambda$.  For our
calculation, this takes the form of a photon squared-mass,  
$\lambda \equiv m_\gamma^2$.  Let 
${\cal A}$ be the amplitude for a generic process involving charged
particles.  To all orders in the EM interaction ${\cal A}$ is given by the
expansion
\beq
{\cal A} \ = \ \sum_{k=0}^\infty \ {\cal A}_k \ \ , 
\label{cap0e6}
\eeq
with ${\cal A}_k = {\cal O}(\alpha^k)$.   
Order by order one has the sequence of 
relations,
\beqa
{\cal A}_0 &=& a_0 \ \ , 
\nonumber \\
{\cal A}_1 &=& a_0 \cdot \alpha B (\lambda) + a_1  \ \ ,
\nonumber \\
{\cal A}_2 &=& a_0 \cdot {(\alpha B (\lambda))^2 \over 2} 
 + a_1  \cdot \alpha B(\lambda) + a_2  \ \ ,
\label{cap0e7} \\
&\vdots & 
\nonumber 
\eeqa
Here $B(\lambda)$ is an infrared divergent function of $\lambda$, 
while the $a_k$ are infrared finite.  Summing to all
orders results in the compact relation
\beq
{\cal A}(\alpha) \ = \ e^{\alpha B (\lambda)} \cdot 
\sum_{k = 0}^\infty \ a_k  \ \equiv 
e^{\alpha B (\lambda)} {\bar {\cal A}}(\alpha) \ \ , 
\label{cap0e8}
\eeq
where all IR-singular dependence appears solely within the
exponentiated factor $\alpha B (\lambda)$, which multiplies the infrared
finite amplitude ${\bar {\cal A}}(\alpha)$.  The function $\alpha B
(\lambda)$ only depends on the external states and knows nothing on
the nature of the interaction generating the process.  On the other
hand, the amplitude ${\bar {\cal A}}(\alpha)$ contains the structure
dependent EM effects.  These quantities  arise naturally in the
calculation described in Ref.~\cite{em2}, where we  
provided explicit expressions in the case of ${\cal A}_{+-}$. 

The construction just described needs to be supplemented by the the
following comments.  As Eq.~(\ref{cap0e7}) shows, the function $B
(\lambda)$ arises as a first order correction in $\alpha$. 
This means that a one-loop calculation allows resummation of the IR
singularity to all orders.  However, the definition of the IR finite
part of $B(\lambda)$ is not unique. This means that there is an
ambiguity in the way one separates the infrared multiplicative factor
from the structure dependent effects. This is a peculiar  property of
EM radiative corrections and does not affect the definition of
physical observables.  For example, once one picks a definition for
$B(\lambda)$ and follows it throughout the calculation, 
comparison with experiment will lead to unambiguous extraction of the 
EM free quantities (like $a_0$). We will explicitly display our
formulas for $B(\lambda)$ below.

The infrared divergences of the amplitudes have now been isolated
in an overall factor.  Removal of infrared divergences from the
expression for the decay rate or cross sections is achieved by taking
into account the effect of soft real photons in the external states.
This is motivated by the observation that for soft photons, whose
energy is below some experimental resolution $\omega$, the generic states 
% $\pi\pi$ and $\pi \pi \gamma$ 
$n$ and $ n + k \, \gamma$ cannot be distinguished. The physical
observable always involves an inclusive sum over the $n$ and $n + k \,
\gamma$ final states. We shall give details of this in Sect.~\ref{fit}.

\subsection{Isospin Amplitudes} 

Having described the construction of IR finite amplitudes, we can now
analyze the effect produced on the isospin amplitudes $A_0, A_2$.  We
start from the IR finite amplitudes in the charged basis $\{
\overline{\cal A}_{+-}, \overline{\cal A}_{00}, \overline{\cal A}_{+0}
\}$. It is then possible to define the would-be isospin amplitudes by
taking the following linear combinations:
\beqa \overline{\cal A}_{0} & = & \frac{2}{3} \overline{\cal A}_{+-} +
\frac{1}{3} \overline{\cal A}_{00} \ \ , \nonumber \\ \overline{\cal A}_{2} 
& = & \frac{\sqrt{2}}{3} ( \overline{\cal A}_{+-} - \overline{\cal A}_{00}
) \ \ ,
\label{cap0e12}  \\ 
\overline{\cal A}_{2}^{+} & = & \frac{2}{3} \overline{\cal A}_{+0} \ \
. \nonumber \eeqa
The content of Eq.~(\ref{cap0e12}) is that in the absence of EM and for
$m_u = m_d$ the amplitudes $\overline{\cal A}_I$ truly describe
transitions to $\pi \pi$ states with definite isospin. In the presence of 
isospin breaking interactions, however, the amplitudes 
$\overline{\cal A}_I$ become 
\beq \overline{\cal A}_{I} \ = 
 \left( A_I + \delta A_I \right) \, e^{i (\delta_I +
\gamma_I)} \ ,
\label{cap0e14}
\eeq
with shifts $\delta A_I$ and $\gamma_I$ corresponding 
respectively to the modulus and
phase of the original isospin amplitude.

We can summarize the discussion presented so far by displaying the 
parameterization of the IR finite amplitudes in the charged basis 
\beqa
\overline{\cal A}_{+-} &=& \left( A_{0} + \delta A_0 \right)
e^{i(\delta_0 + \gamma_0)} + { 1 \over \sqrt{2}} \left( A_{2} +
\delta A_2  \right)
e^{i(\delta_2 + \gamma_2)}
\ , \nonumber \\
\overline{\cal A}_{00} &=& \left( A_{0} + \delta A_0  \right)
e^{i(\delta_0 + \gamma_0)} - \sqrt{2} \left( A_{2} +
\delta A_2  \right)
e^{i(\delta_2 + \gamma_2)} \ \ ,
\label{cap5e4} \\ 
\overline{\cal A}_{+0} &=& {3 \over 2} \left( A_{2}
+ \delta A_2^{+}  \right)
e^{i (\delta_2 + \gamma_2^{+})} \ \ ,
\nonumber
\eeqa
where $\delta A_{I}$ and $\gamma_I$ contain IR finite EM effects as well as
other isospin breaking terms.  This parameterization has to be
compared with the isospin invariant expressions in Eq.~(\ref{cap0e2}).
We recall here that this parameterization holds for the IR finite
amplitude as defined in Eq.~(\ref{cap0e8}).  We also observe that the 
shifts $\delta A_2^{+}$ and $\gamma_2^+$ in $\overline{\cal A}_{+0}$
are distinct from the corresponding shifts in $\overline{\cal A}_{+-}$
and $\overline{\cal A}_{00}$, as a consequence of the $\Delta I = 5/2$
amplitude induced by electromagnetism.  In the notation of
Eq.~(\ref{cap0e4}) one has:
\begin{eqnarray}
 \delta A_{0} & = &  \delta A_{1/2}  \ ,  \nonumber \\
 \delta A_{2} & = &   \delta A_{3/2}  +  A_{5/2} \ ,  \label{cap5e4.1} \\
 \delta A_{2}^{+} & = & \delta A_{3/2}  -  2/3 \ A_{5/2} \ 
 . \nonumber  
\end{eqnarray}

\subsection{Discussion}

The parameterization given in Eqs.~(\ref{cap5e4}),(\ref{cap5e4.1})
provides the basis for any phenomenological analysis of $K \rightarrow
\pi \pi$ decays with inclusion of isospin breaking (due to strong and
electromagnetic effects).  Comparison with experimental branching
ratios (see next Section for some caveats related to radiative
corrections) allows one to arrive at relations among the different
parameters. Examples of such analysis are given in Ref.~\cite{valencia} 
and the third paper of Ref.~\cite{isobnew}. 

It is legitimate at this point to ask what do we know about the
parameters entering Eq.~(\ref{cap5e4}) and what do we want to 
extract from the comparison with experiment. 
\begin{enumerate}
\item First, let us consider the amplitudes $A_{0}$ and $A_{2}$. 
Many theoretical efforts have been devoted to their calculation, and 
no calculation can be considered fully satisfactory at present. 
We want to extract these parameters from the comparison with the branching 
ratios, eliminating the electromagnetic isospin breaking contaminations. 
\item Next, one has the shifts $\delta A_{0}$, $\delta A_{2}$, and
$\delta A_{2}^{+}$. We have calculated the EM contributions to them
\cite{em2,em3}, and we are quite confident that our results capture the
true underlying physics within the theoretical uncertainty quoted.  In
particular, our results provide an estimate for $A_{5/2}$.  Explicitly
we find~\cite{em3}: 
\beqa \delta A_{0}^{\rm EM} & = &  (0.0253 \pm 0.0072) \cdot 
	10^{-7} \,  M_{K^{0}}  \ , \nonumber \\
\delta A_{2}^{\rm EM} & = &  (0.0147 \pm 0.0063) \cdot 
	10^{-7} \,  M_{K^{0}}   \ , \nonumber \\ 
\delta A_{2}^{+ \ \rm EM} & = &  (0.008 \pm 0.0088) \cdot 
	10^{-7} \,  M_{K^{0}}  \ , \nonumber \\ 
A_{5/2}^{\rm EM} & = &  (0.0137 \pm 0.0097) \cdot 
	10^{-7} \,  M_{K^{0}}  \ .  \nonumber 
\eeqa 
For recent estimates of the size of non-EM isospin breaking effects,
see Ref.~\cite{isobnew}.  
The corrections $\delta A_{\Delta I}^{\rm iso-brk}$ contain only a
negligible $\Delta I = 5/2$ component, as the only source for it would
be a combination of the $\Delta I = 1$ interaction proportional to
$(m_u - m_d)$ with the suppressed $\Delta I = 3/2$ weak interaction.
This ensures that $\delta A_{1/2,3/2}^{\rm iso-brk}$ can be
reabsorbed into $A_{0,2}$.  Therefore, in what follows $A_{0}$ and
$A_{2}$ still contain strong isospin breaking contaminations. 
These can be subtracted by using the results of Refs.~\cite{isobnew}. 

\item Finally, let us consider the rescattering phases.  In absence of
isospin-breaking the phases $\gamma_I$ vanish as a consequence of
Watson's theorem. The strong phases $\delta_I$ at $s = M_K^2$ are
known through dispersive treatment of $\pi \pi$ scattering data
\footnote{For a list of references see \cite{valencia}.}.  The
inclusion of strong isospin breaking effects still gives $\gamma_I =
0$, to first order in $m_u - m_d$ \cite{valencia}.  It is the
inclusion of EM corrections that generates nonzero $\gamma_I$.  Since
the phase $\delta_2 + \gamma_{2}^{+}$ does not enter any physical
relation, we disregard it from now on. As for the phases
$\gamma_0$ and $\gamma_2$, we can relate them to EM effects in the
final state interaction. This can be done in perturbation theory
\cite{em2} or in a more general setting provided by unitarity.  As we
shall discuss in Sect.~\ref{phases}, our present knowledge of 
$\gamma_{0,2}$ reveals a large value of $\gamma_2$ and relies on a
lowest order analysis of EM corrections to $\pi \pi$ scattering.  Next
to leading order corrections can be quite large at $s = M_K^2$, and
this adds substantial uncertainty to $\gamma_2$.
\end{enumerate}

\section{Analysis in the Presence of Electromagnetism}
\label{full}

The CP-conserving sector of $K \rightarrow \pi \pi$ phenomenology
relies essentially on three experimental numbers. These are the
partial decay widths of $K_S$ into $\pi^+ \pi^- $, $\pi^0 \pi^0$
and of $K^{\pm}$ into $\pi^{\pm} \pi^0$. Knowledge of these numbers
allows one to extract the invariant decay amplitudes and to compare
with theoretical calculations.  In this section we
describe the procedure to be followed in order to extract the invariant
amplitudes in presence of radiative corrections. 

\subsection{The Fitting Procedure in Presence of Radiative Corrections}

In presence of radiative corrections, the appropriate expression to be 
used when comparing theory and experiment is: 
\beq
\Gamma_{n} (\omega) \ = \ \frac{\Phi_{n}}{2 \, \sqrt{s_{n}}} \, 
	| \overline{\cal A}_{n} |^2 \, G_{n} (\omega) \ \ ,
\label{cap5e1}
\eeq
where $n = \{ +-, 00, +0 \} $. In Eq.~(\ref{cap5e1}) the left hand
side $\Gamma_{n}$ represents the measured partial width and we
indicate with the parameter $\omega$ its dependence on the way soft
photons are treated in the data analysis. It is clear that the use of
different cuts leads to different values for the decay widths, because
of the inclusion of different portions of the corresponding radiative
channel ($n + \gamma$) in the data sample.  On the right hand side of
Eq.~(\ref{cap5e1}) one has the kinematical parameters $s_n$ (squared
center-of-mass energy of the process) and $\Phi_{n}$ (the two body
invariant phase space associated with the decay).  The quantities
related to the dynamics are $\overline{A}_{n}$ and $G_{n} (\omega)$.
$\overline{A}_{n}$ is the infrared finite invariant amplitude, as
defined in Sect.~\ref{IRB}, Eq.~(\ref{cap5e4}). It contains the true
weak transition component that we wish to ultimately extract as well
as infrared finite electromagnetic corrections.  $G_{n} (\omega)$ is
the infrared factor associated with the combined effect of virtual and
real photons. The latter contribution involves an integration over the
soft-photon phase space: this has to be done with the {\it same}
prescription used in extracting the experimental number $\Gamma_{n}
(\omega)$.

The extraction of IR-free invariant amplitudes is a straightforward
consequence of Eq.~(\ref{cap5e1}).  Specializing to the $K \rightarrow
\pi \pi$ case, one has:
\beqa
\frac{| \overline{\cal A}_{+-} |}{M_{K^0}} \ & = & \ \frac{1}{\sqrt{2}} \, 
	\sqrt{   \frac{ 8 \pi}{p_{+-}} \, \frac{ \Gamma_{+-} (\omega)}{
	G_{+-} (\omega)} } \ , \nonumber \\
\frac{| \overline{\cal A}_{00} |}{M_{K^0}} \ & = & \ 
	\sqrt{   \frac{ 8 \pi}{p_{00}} \, \Gamma_{00}  } \ , 
\label{cap5e2}   \\
\frac{| \overline{\cal A}_{+0} |}{M_{K^+}} \ & = & \ 
	\sqrt{   \frac{ 8 \pi}{p_{+0}} \, \frac{ \Gamma_{+0} (\omega)}{
	G_{+0} (\omega)} } \ , \nonumber 
\eeqa
where 
\beqa
p_{+-} & = & \sqrt{ \left(\frac{M_{K^0}}{2} \right)^2 - M_{\pi^{+}}^2 } \ , 
 \nonumber \\
p_{00} & = & \sqrt{ \left(\frac{M_{K^0}}{2} \right)^2 - M_{\pi^{0}}^2 } 
	 \ \ , \\
p_{+0} & = & \sqrt{ \left(\frac{M_{K^+}}{2} + \frac{M_{\pi^0}^2 - 
M_{\pi^+}^2}{2 M_{K^+}} \right)^2 - M_{\pi^{0}}^2 } \  \ .   \nonumber 
\label{cap5e3}
\eeqa 
In order to carry out the procedure one needs the physical masses, the
experimental input $\Gamma_{n} (\omega)$, and the corresponding
theoretical infrared parameters $G_{+-,+0} (\omega)$. If things are
done properly, the $\omega$ dependence cancels in the ratios on the
right hand side of Eq.~(\ref{cap5e2}), which provides then the values
of $|\overline{\cal A}_n|$.  Having the amplitudes $|\overline{\cal
A}_n|$ one can then use Eq.~(\ref{cap5e4}) to obtain relations between
the physical parameters.  However, there is an important issue to be
addressed in order to complete the program outlined above: it is the
proper definition of the infrared factors $G_{+-}(\omega)$ and
$G_{+0}(\omega)$, to which we now turn.

\subsection{The Infrared Factors} 
\label{Infsing}

As already noted, the theoretical definition of $G_{+-}(\omega)$ and
$G_{+0}(\omega)$ involves integration over the soft-photon phase
space, with cuts generically indicated by $\omega$. 
The
infrared factor {\em is} experiment dependent and accounts for the
component of the radiative mode ($\pi \pi \gamma$ in our case)
included in the parent mode ($ \pi \pi$) branching ratio.
We shall work to order $\alpha$ and thus include only the
effect of the $\pi \pi \gamma$ radiative state. 
%(see Eq.~(\ref{cap0e11}) ) 
While we display the ingredients needed for any experimental analysis,
we carry the treatment to completion for the case where the 
infrared sensitivity is isotropic in the center of mass. 
That is, we display the explicit formulas obtained when one
integrates over the $\pi \pi \gamma$ phase space applying an
isotropic cutoff $\omega$ on the photon energy $E_{\gamma}$ in the
center of mass frame.   
This is most natural to use in the data analysis
for an experiment such as KLOE at DA$\Phi$NE, given the working conditions of
the machine and the detector geometry.  A more detailed study (either
theoretical or experimental) is required in order to apply radiative 
corrections to the data analysis of other experiments.  The presence 
of several high statistics experiments offers a unique opportunity for 
performing an accurate measurement of $K \rightarrow
\pi \pi$ branching ratios, including the effect of radiative
corrections.  Such an analysis would fill the present gap in the
study of radiative corrections to $K^{0} \rightarrow
\pi \pi$, and would therefore be highly desirable.

Calculation of the factors $G_{+-}(\omega)$ and $G_{+0}(\omega)$
requires consideration of a 
combination of effects due to virtual photons in the
amplitudes ${\cal A}_{+-, +0}$ (field theoretic version of the
non-relativistic Gamow factors) and to soft real photons entering the 
process $K \to \pi \pi + \gamma$.
Moreover, in the infrared region of the spectrum,  the amplitude for $K
\to \pi \pi + \gamma$ is dominated by the internal bremsstrahlung (IB)
component, proportional to the non-radiative amplitude. For the case in
question one has:
\beqa
{\cal A}^{IB}_{+ - \, \gamma} & = & e {\cal A}_{+-}  \, \left(
\frac{ \epsilon \cdot  p_{+}}{ q \cdot p_{+}}
 - \frac{ \epsilon \cdot  p_{-}}{ q \cdot p_{-}} \right)  \ ,
 \nonumber \\
{\cal A}^{IB}_{+ 0 \, \gamma} & = & e {\cal A}_{+0}  \, \left(
\frac{ \epsilon \cdot  p_{+}}{ q \cdot p_{+}}
 - \frac{ \epsilon \cdot  p_{K}}{ q \cdot p_{K}} \right)  \ ,
\label{cap5e5}
\eeqa
where $\epsilon$ and $q$ are the polarization and momentum of the
emitted photon. Now, the infrared-finite observable decay
rate is
\begin{equation}
\left. \Gamma_{n} ( \omega ) =  \Gamma_{n}  \  +  \
\Gamma_{n \, \gamma} (\omega) \ \ \ \ ,
  \right.
\label{cap5e6}
\end{equation}
where $n = +-, +0$ and 
\begin{eqnarray}
\Gamma_{n} & = & \frac{1}{2 M_{K}} \, \int \, d \Phi_{n} \
 \big| \overline{\cal A}_{n} \, \left( 1 + \alpha B_{n} (m_{\gamma}) 
\right) \,  \big|^{2} \ \ , 
 \label{cap5e6.5} \\
\Gamma_{n \, \gamma} (\omega) & = & \frac{1}{2 M_{K}} \,
\int_{E_{\gamma} < \omega} \, d \Phi_{n \gamma}  \
 \big| {\cal A}_{n \gamma} \big|^{2}  
=  \frac{1}{2 M_{K}} \,  \int \, d \Phi_{n} \,
   \big| \overline{\cal A}_{n} \big|^2 \, \ {I}_{n}(m_{\gamma},\omega) \ \ . 
\label{cap5e7}
\end{eqnarray}
At this stage the rest of the calculation becomes dependent on the 
geometry of the experiment. For a kaon at rest with an isotropic detector,
the acceptance cutoff $E_\gamma < \omega$ will be the same in all directions. 
For a kaon in flight, the acceptance may be different for photons emitted
in different directions. In this latter situation, the integral in 
Eq.~(\ref{cap5e7}) needs to be numerically integrated over the detector
acceptance and then added to the two body result,
Eq.~(\ref{cap5e6.5}).  This will then be finite in the limit of 
$m_\gamma \to 0$, and will allow the measurement of the IR 
finite amplitude $\overline{\cal A}_{n}$. We carry out this 
procedure explicitly below for the case of an isotropic cutoff. 
 
In Eqs.~(\ref{cap5e6.5}),(\ref{cap5e7}) $d \Phi_{k}$ is the differential
phase space factor for each process, $\overline{\cal A}_{n}$ is the IR
finite amplitude, as defined in Sect.~\ref{IRB} by extracting the
IR divergent functions $B_{n}$.  The $B_{n}$ can be calculated by
considering one-loop diagrams with virtual photons connecting the
external legs and using a point-like vertex for the weak
interaction. The definition of $B_n$ is not unique due to the
possibility of adding IR and UV finite terms to it. The explicit
expressions given below \footnote{See also Ref.~\cite{em2}.}
fully specify our choice of what goes into $B_{n}$ and what goes into the
structure dependent shifts $\delta A_{n}$.  The second expression in 
Eq.~(\ref{cap5e7}) describes the factorization of the $\pi \pi$ and
$\gamma$ phase spaces, valid with an accuracy of 
$\omega/M_K$. Explicitly one has:
\beq
I_{n} (m_{\gamma}, \omega) =  \int_{E_{\gamma} < \omega} \, 
\frac{d^3 q}{(2 \pi)^3 2 E_{\gamma}} \, \sum_{pol} \, 
\bigg| \frac{{\cal A}^{IB}_{n \gamma} }{ {\cal A}_{n} } \bigg| ^2  \ . 
\label{cap5e7.5}
\eeq
Combining the above results one arrives at 
\beq
G_{n} (\omega) =  1 + 2 \alpha \, {\cal R}e {B}_{n} (m_{\gamma}) \,
+ \, I_{n} \, (m_{\gamma},\omega) + {\cal O} (\alpha^2)  \ \ .
\label{cap5e8}
\eeq

We now collect the explicit form of the 
functions $B_{+-,+0}$ and $I_{+-,+0}$, entering in the definition 
of  $G_{+-,+0}$  (see Eq.~\ref{cap5e8}). For $G_{+-}(\omega)$ 
one needs: 
\beqa
{B}_{+-} ( m_\gamma^2) &=& {1 \over 4 \pi} \bigg[
2 a (\beta) \ln {M_\pi^2 \over m_\gamma^2} + H_{+-} (\beta) 
+ i \pi \left( {1 + \beta^2 \over \beta}
\ln { M_K^2 \beta^2 \over m_\gamma^2} - \beta \right) \bigg] \ \ ,
\nonumber \\
{I}_{+-}( m_\gamma , \omega ) & = & {\alpha \over \pi} \left[
a (\beta) \ln \left( { m_\gamma \over 2 \omega } \right)^2 +
F_{+-} ( \beta ) \right]  \ , \label{cap5e9} 
\eeqa
where
\beq
\beta = (1 - 4 M_\pi^2 /M_K^2 ) ^{1/2}
\label{cap5e10}
\eeq
and
\beqa
a (\beta) &=& 1 + {1 + \beta^2 \over 2 \beta}
\ln \left( {1 - \beta \over 1 + \beta} \right) \ \ ,
\nonumber \\
H_{+-} ( \beta) & = & 
{1 + \beta^2 \over 2 \beta} 
\left[  \pi^2 + \ln {1 + \beta \over 1 - \beta}
\ln {1 - \beta^2  \over 4 \beta^2 } + 2 f \left(
{1 + \beta \over 2 \beta} \right) - 2 f
\left( { \beta - 1 \over 2 \beta} \right) \right] 
\nonumber \\
& \ \   &  + \, 2 \, + \,  \beta \ln {1 + \beta \over 1 - \beta} \ \ ,   
\nonumber  \\
 F_{+-} ( \beta ) & = &  {1 \over \beta} \ln {1 + \beta \over 1 - \beta}
+ {1 + \beta^2 \over 2 \beta} \bigg[
2 f ( -\beta) - 2 f (\beta) + f \left({1 + \beta \over 2}\right)
\nonumber \\
& & \phantom{xxxxx} - f \left({1 - \beta \over 2}\right) + {1 \over 2}
\ln {1 + \beta \over 1 - \beta} ~\ln (1 - \beta^2 ) +
\ln 2 ~\ln {1 - \beta \over 1 + \beta} \bigg] \nonumber \\
f(x) &=& - \int_0^x dt ~ {1\over t} \ln|1 - t| \ \ .  
\label{cap5e11} 
\eeqa
We note that $B_{+-}$ includes the Coulomb factor $\pi \alpha /v_{\rm rel}$ 
(first term in $H_{+-} (\beta)$), as well as typical field theoretic 
effects.   As for $G_{+0} (\omega)$, one has:  
\beqa
{B}_{+0} ( m_\gamma^2) & = & {1 \over 4 \pi} \left[
2 b (\beta) \ln {M_\pi^2 \over m_\gamma^2} + H_{+0} (\beta) \right] \ , 
\nonumber \\
{I}_{+0}( m_\gamma , \omega ) & = & {\alpha \over \pi} \left[
b (\beta) \ln \left( { m_\gamma \over 2 \omega } \right)^2 +
F_{+0} ( \beta ) \right]   \ ,
\label{cap5e12}
\eeqa
where 
\beqa
b (\beta) &=& 1 + \frac{1}{2 \beta} \, \log \left( 
\frac{ 1 - \beta}{ 1 + \beta} \right) 
 \ \ , \nonumber \\
H_{+0} ( \beta) & = & - \frac{1}{\beta} \left[ 
\frac{1}{2} \log \left( \frac{1 + \beta}{2} \right)  \log 
\left( \frac{ 2 + 2 \beta }{ (1 - \beta)^2} \right) \right. 
\nonumber \\
& & \left.  - \frac{1}{2} 
\log \left( \frac{1 - \beta}{2} \right)   
\log \left( \frac{ 2 -  2 \beta }{ (1 + \beta)^2} \right) \right. 
 \nonumber  \\
&   & \left. + \log \left( \frac{4 \beta}{1 - \beta^2} \right) \, 
\log \left( \frac{ 1 + \beta}{1 - \beta} \right)  + 
f \left( \frac{1 + \beta}{2 \beta} \right) - 
f \left( \frac{\beta - 1}{2 \beta} \right)  \right. 
\nonumber \\
&  &  + \left. f \left( - \frac{(1 -\beta)^2}{4 \beta} \right) - 
f \left(  \frac{(1 + \beta)^2}{4 \beta} \right) \right]
\nonumber \\
&  & + 2 + \log \frac{1 - \beta^2}{4} - 
\frac{2}{1 - \beta} \log \left( \frac{1 + \beta}{2} \right) -
\frac{2}{1 + \beta} \log \left( \frac{1 - \beta}{2} \right)  \nonumber 
\label{cap5e13} 
\eeqa
and 
\beqa
F_{+0} ( \beta ) & = &  1 + \frac{1}{2 \beta} \, \log \left(
\frac{1 + \beta}{1 - \beta} \right) 
\nonumber \\
& & - \frac{4}{1 - \beta^2} \, \int_{-1}^{+1} 
	\, dx \frac{E (x)}{D(x) p(x)} \, 
\log \left( \frac{E(x) + p (x)}{E(x) - p (x)} \right) \, 
\nonumber \\
D(x) & = & (x - x_1) (x - x_2) \nonumber \\
x_{1/2} & = & \frac{M_{K}^{2}}{M_{\pi}^{2}} (1 \pm \beta) - 1 \nonumber \\
E(x) & = & \frac{M_{K}}{4} (3 - x)   \nonumber \\
p(x) & = & \frac{M_{K}}{4} \beta (1 + x)  \ \ . 
\label{cap5e14} 
\eeqa

\begin{figure}[htb]
\centering
\begin{picture}(200,150)  
\put(50,20){\makebox(100,120){\epsfig{figure=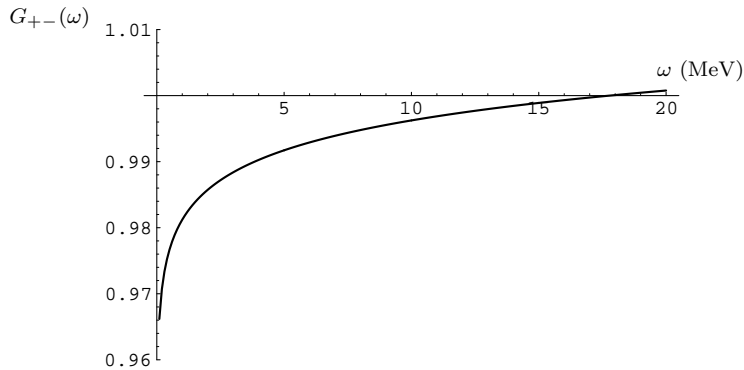,height=3in}}}
\put(200,125){\scriptsize{$\omega$ (MeV)}}
\put(-45,145){\scriptsize{$G_{+-} (\omega)$}}
\end{picture}
\caption{The function $G_{+-} (\omega)$.\hfill 
\label{cap5fig1}}
\end{figure}

\begin{figure}[htb]
\centering
\begin{picture}(200,150)  
\put(50,20){\makebox(100,120){\epsfig{figure=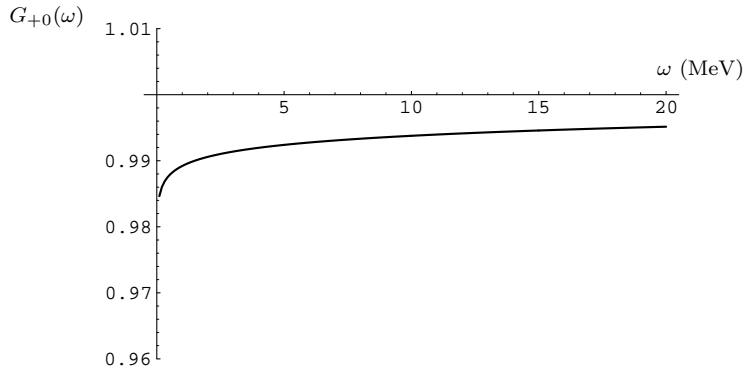,height=3in}}}
\put(200,125){\scriptsize{$\omega$ (MeV)}}
\put(-45,145){\scriptsize{$G_{+0} (\omega)$}} 
\end{picture}
\caption{The function $G_{+0} (\omega)$.\hfill 
\label{cap5fig2}}
\end{figure}

$G_{+-}$ and $G_{+0}$ do not depend on the infrared regulator
$m_{\gamma}$.  We display plots of the functions $G_{+-}(\omega)$ and
$G_{+0}(\omega)$ on a typical range of values for the parameter
$\omega$ in Figs.~\ref{cap5fig1} and \ref{cap5fig2}.  The functions
$G_{n}(\omega)$ are the only EM effects previously considered in the
literature~\cite{old}, although with a slightly different definition.
In fact, the works of Ref.~\cite{old} use a point-like vertex for the
weak interaction, and therefore are not sensitive to structure
dependent corrections. In these works the EM effects due to
wavefunction renormalization and vertex correction go entirely in the
definition of $B_{n}$ (this includes also the UV divergent terms,
regulated by means of a cutoff).  Apart from the cutoff-dependent term
and an extra finite contribution, our expressions match the ones given
in the second and third papers of Ref.~\cite{old}.
\section{Effect of EM on $K \rightarrow \pi \pi$ Phases}
\label{phases}

The amplitude parameterization we have used for the previous analysis
already implies that $K \rightarrow \pi \pi$ data do not provide direct 
information on the strong $\pi \pi$ phase shift difference $\delta_0 -
\delta_2$.  This would only be true in the isospin limit ($\gamma_I =
0$). In this limit, the requirement that strong interaction phases
appear in the weak decays is known as Watson's theorem, which is valid
whenever the final state rescattering involves only elastic
scattering.  Despite the fact that isospin breaking has been long 
understood to cause mixing of weak amplitudes, 
there has been no recognition that the strong interaction phases no
longer suffice to describe the rescattering effects. This occurs
because elastic rescattering is no longer the full content of final
state interaction, so that the conditions for the application of
Watson's theorem no longer apply. Moreover, there exists a sizeable 
discrepancy between the determination of $\delta_0
- \delta_2$ from $K \rightarrow \pi \pi$ data using isospin relations
and the favored value of the phase shift difference known by other 
determinations. This
seems to point to violations of Watson's theorem.  It is our purpose
to set the framework for the correct treatment of this problem in the
isospin breaking real world. We shall accomplish this by writing a
coupled channel unitarity constraint in the presence of EM interactions
(and isospin breaking in general) and solving for the parameters
$\gamma_0$ and $\gamma_2$ entering Eq.~(\ref{cap5e4}).  This analysis 
will complement and extend the perturbative results obtained in 
Ref.~\cite{em2}, which already indicated a large value of $\gamma_2$. 
We defer to the next section the extraction of $\delta_0 - \delta_2$ 
and the uncertainty to be associated with it. 

\subsection{Extended Unitarity Relations} 

The first step in our program is writing down meaningful unitarity
relations in the presence of EM interactions. Here, it is
natural to work in the {\em charged} basis $\{ \pi^+ \pi^-, \pi^0
\pi^0 \}$ and then try to recover the notion of isospin amplitudes. 
In order to fix the notation, let us start from the unitarity
relations involving the decay amplitudes of $K^0$ to $\{ \pi^+ \pi^-,
\pi^0 \pi^0 \}$ in the limit in which EM is turned off.  Then, 
only the $\pi\pi$ intermediate states have to be taken into account
and one finds:
\beqa
{\cal A}_{+-} - {\cal A}_{+-}^* &=& i \left(
{\cal T}_{+-;+-}^* \times {\cal A}_{+-} +
{\cal T}_{00;+-}^* \times {\cal A}_{00} \right) \ \ ,
\nonumber \\
{\cal A}_{00} - {\cal A}_{00}^* &=& i \left(
{\cal T}_{+-;00}^* \times {\cal A}_{+-} +
{\cal T}_{00;00}^* \times {\cal A}_{00} \right) \ \ ,
\label{cap5e17} 
%{\cal A}_{+0} - {\cal A}_{+0}^* &=& i
%{\cal T}_{+0;+0}^* \times {\cal A}_{+0} \ \ .
%\nonumber
\eeqa
In Eq.~(\ref{cap5e17}) ${\cal A}_{+ - }$ and ${\cal A}_{00} $
represent the $K^0$ decay amplitudes and ${\cal T}_{f;i}$ is the
$T$-matrix element for the transition $ i \rightarrow f$ ( in this
case it only involves pion-pion scattering ).  `${\cal T}^* \times
{\cal A}$' denotes the product of amplitudes integrated
over the intermediate state phase space.  In the case considered here
of two-pion intermediate states, one has: 
\beq
{\cal T}^* \times {\cal A} \ \equiv \ \int d\Phi_2  \
{\cal T}^* {\cal A} \ =   \Phi_s  \, \ 4 \beta ~ {\cal A}
\cdot {\cal T}^{*} \ , 
\label{cap5e18}
\eeq 
where $\beta = (1 - 4 m_\pi^2 /m_K^2)^{-1/2}$ is the pion
velocity in the kaon rest frame. $\Phi_s $ is the symmetry factor for
identical particles (equal to $1/2$ for the $\pi^0 \pi^0$ state) and 
${\cal T}$ is the S-wave projection of the $\pi \pi$ scattering amplitude 
$T (\cos \theta)$, defined by:
\beq
{\cal T} = \frac{1}{64 \pi} \, \int_{-1}^{+1} \, d (\cos \theta)  \, 
T (\cos \theta)  \ . 
\eeq
%
%MAYBE HERE WRITE SOMETHING ON HOW ONE WOULD PROCEDE IN THE ISOSPIN 
%LIMIT 

Turning on EM interactions introduces isospin breaking dynamics as
well as IR singularities in the amplitudes and the opening of
intermediate radiative channels.  Specifically, ${\cal A}_{+ - }$,
${\cal T}_{00, +- }$, and ${\cal T}_{+-, +-}$ become IR divergent and
${\cal T}_{+-, +-}$ acquires a purely Coulomb component (also IR
singular).  The work of Refs.~\cite{F61,Weinberg65}, summarized in
Sect.~\ref{IRB} teaches us that one can always isolate the singularity
in a multiplicative exponential factor 
\beq
{\cal A}_{f,i}  \ = \ e^{\alpha B_{f,i}} \  \overline{\cal A}_{f,i}   \  .   
\label{cap5e19}
\eeq
Here $\overline{\cal A}_{f,i} $ is the IR finite amplitude and
$B_{f,i} (m_{\gamma})$ is the IR singular factor that depends only on
the external states. We shall only need the factor $B_{+-}$, already
encountered in this paper, associated
with a pair of charged pions in the initial or final state.  We note
that $B_{f,i} (m_{\gamma})$ is in general complex.  In particular, its
imaginary part is equal to the Coulomb scattering phase shift
associated with each pair of charged particles in the initial and
final states~\cite{Weinberg65}.

Upon integrating over the phase space and using the above mentioned
property on the Coulomb phases, one can rewrite Eq.~(\ref{cap5e17}) in
terms of IR finite quantities and $B_{f,i}$ factors.  Moreover, due to
the integration over the phase space some contributions in the
%simplifies the structure of the
$B_{f,i}$ factors simplify and one ends up with:
\beqa
\overline{\cal A}_{+-} -  \overline{\cal A}_{+-}^* &=& i \left(
\overline{\cal T}_{+-;+-}^* \times  \overline{\cal A}_{+-}  \,  e^{2 \alpha {\cal R}e B_{+-} }  +
\overline{\cal T}_{00;+-}^* \times  \overline{\cal A}_{00} \right) \ \ ,
\nonumber \\
\overline{\cal A}_{00} - \overline{\cal A}_{00}^* &=& i \left(
\overline{\cal T}_{+-;00}^* \times  \overline{\cal A}_{+-}  \,   e^{2 \alpha {\cal R}e B_{+-} }   +
\overline{\cal T}_{00;00}^* \times \overline{\cal A}_{00} \right) \ \ . 
\label{cap5e20}
\eeqa
Note that now $\overline{\cal T}_{+-, +-}$ is the IR finite $\pi^+
\pi^- \rightarrow \pi^+ \pi^- $ amplitude subtracted of its purely
Coulomb term.  

Eq.~(\ref{cap5e20}) contains IR singularities, but the
analysis is still missing an important effect of EM: the opening of
inelastic radiative channels.  This is the key ingredient in obtaining
an IR finite set of unitarity constraints, as it was in obtaining an IR
finite cross section or decay rate.  In fact, the IR singularities
will cancel in the sum over the $\pi^+ \pi^-$ and $\pi^+ \pi^- \gamma$
intermediate states, with the same mechanism described in the
definition of $\Gamma_{+-}, \Gamma_{+0}$ earlier in Sect.~\ref{full}.   
Working at order ${\cal O} (\alpha)$, we consider
only the radiative state $\pi^+ \pi^- \gamma$. For our analysis
we require the amplitudes for $K^0 \rightarrow \pi^+ \pi^- \gamma$ and
$\pi \pi \rightarrow \pi^+ \pi^- \gamma$. We include only the
internal bremsstrahlung component of these amplitudes, known to be
dominant over possible direct emission terms.  Now one has to
integrate over the full $\pi^+ \pi^- \gamma$ phase space and the final
result for the unitarity condition reads:
\beq
 {\cal I}m~ {\overline{\cal A}_{+-} \choose
\overline{\cal A}_{00} } =  \beta 
\left( \begin{array}{ll}
2 \, \overline{\cal T}_{+-;+-}^{*} \left( 1 + \Delta_{+-} \right) \ \ &
\ \ \ \overline{\cal T}_{00;+-}^* \\
2 \, \overline{\cal T}_{+-;00}^*  \left( 1 + \Delta_{+ -} \right) \ \  &
\ \ \ \overline{\cal T}_{00;00}^*
\end{array}\right)
{ \overline{\cal A}_{+-} \choose
\overline{\cal A}_{00} } \ \ .
\label{cap5e21}
\eeq
We recall that $\overline{\cal T}_{a,b} $ are the S-wave projections
of the $\pi \pi$ scattering matrix.
$\Delta_{+-}$ is the IR finite remnant of the sum of IR
singular terms in the $\pi^+ \pi^-$ and $\pi^+ \pi^- \gamma$
intermediate states. In terms of the notation of Sect.~\ref{Infsing},
it is given by:
\beq 
\left.
\Delta_{+-} =  - \frac{2 \delta M_{\pi}^{2} }{\beta^2 M_{K}^{2} }    +   
       2 \alpha {\cal R}e  B_{+-} (m_{\gamma})  
+ e^2 \frac{1}{\Phi_{+-}} \, \int d \Phi_{+- \gamma} \ 
\sum_{pol} \bigg| \frac{q_{+} \cdot 
\epsilon}{q_{+} \cdot k} - \frac{q_{-} \cdot \epsilon}{q_{-} \cdot k} 
\bigg|^{2}
 \right. \ . 
\label{cap5e22}
\eeq
%
%e^2 \int_{E_{\gamma} \leq E_D} \frac{d^3 k }{2 (2 \pi)^3 E_{\gamma}}  \ . 
%\frac{\beta (k)}{\beta (0)}  \, \left[ 
Here the first term is the phase space correction due to the EM 
mass-shift of charged pions.  The second term is the effect of infrared
virtual photons, while the third term is the effect of real soft 
photons in the intermediate state $\pi^+ \pi^- \gamma$.  Numerically
we find (displaying separately the phase space contribution and 
the remainder):
\beq
\Delta_{+- } =  \left( - 14.8  +  10.8  \right) \cdot 10^{-3}  = 
 - 4.0 \cdot 10^{-3}  \ \ . 
\label{cap5e23}
\eeq

\subsection{From Charge to Isospin Basis} 

Assuming unitarity of the S matrix, we have thus far obtained a set of
relations containing the IR finite amplitudes in the charge basis. In
order to compare with usual treatments of this problem, we rotate now
to the isospin basis for the $K \rightarrow \pi \pi$ amplitudes,
\beq 
\overline{\cal A}_{\rm ISO} = { \overline{\cal A}_{0} \choose
\overline{\cal A}_{2} } = {1 \over 3} \left( \begin{array}{ll} \ 2 & \
\ 1 \\ \sqrt{2} & - \sqrt{2}
\end{array}\right)
{ \overline{\cal A}_{+-} \choose
\overline{\cal A}_{00} } \ \ . 
\label{cap5e24}
\eeq
Applying the same transformation to the whole system in
Eq.~(\ref{cap5e21}) one obtains in matrix form:
\beq
{\cal I}m \overline{\cal A}_{\rm ISO}  =  \beta \, \left( 
\overline{\cal T}_{\rm ISO}^\dagger \, 
+  {\cal R} \, \right) \overline{\cal A}_{\rm ISO} \ \ , 
\label{cap5e25}
\eeq
where 
\beq
\overline{\cal T}_{\rm ISO} =  
\left( \begin{array}{ll}
T_{0} &  T_{02}\\
T_{20} & T_{2}  
\end{array}\right)  \ , 
\eeq
and 
\beq
{\cal R} = 
\left( \begin{array}{ll}
R_{00} &  R_{02}\\
R_{20} & R_{22}  
\end{array}\right)
=  \frac{1}{3} \, \Delta_{+ - }  \,  
\left( \begin{array}{ll}
2 T_{0}^{*} &    \sqrt{2}  T_{0}^{*} \\
\sqrt{2} T_{2}^{*}  &    T_{2}^{*}  
\end{array}\right)  \ . 
\eeq
$\overline{\cal T}_{\rm ISO}$ is the $\pi \pi$ scattering matrix in the 
isospin basis, while the matrix ${\cal R}$, proportional to 
$\Delta_{+-}$, contains the effect of IR radiative corrections and 
the radiative intermediate channel.  
The $\pi \pi$ scattering T-matrix now involves both strong and EM
interactions, and thus contains isospin-violating matrix elements.  In
the conventions used in our work, the amplitudes for the $\pi \pi$
scattering in the isospin basis are expressed in terms of the charged
ones as:
\beqa 
T_{0} & = & \frac{1}{3} \left( 4 \overline{\cal T}_{+-, +-} +
\overline{\cal T}_{00, 00} + 4 \overline{\cal T}_{00, +-} \right) \ , 
\nonumber \\ 
T_{2} & = & \frac{2}{3} \left( \overline{\cal T}_{+-, +-}
+ \overline{\cal T}_{00, 00} - 2 \overline{\cal T}_{00, +-} \right)  \ , \\
T_{20} = T_{02} & = & \frac{\sqrt{2}}{3} \left( 2 \overline{\cal
T}_{+-, +-} - \overline{\cal T}_{00, 00} - \overline{\cal T}_{00, +-}
\right) \ . \nonumber
\label{cap5e26}
\eeqa
A general parameterization of the $\pi \pi$ transition matrix in 
the isospin basis is:
\beq 
\beta \, \overline{\cal T}_{\rm ISO} = 
\left( \begin{array}{ll} (\eta_0
~e^{2i\delta_0} - 1)/(2 i) & a e^{i(\delta_0 + \delta_2 + \Delta)} \\ a
e^{i(\delta_0 + \delta_2 + \Delta)} & (\eta_2 ~e^{2i\delta_2} -
1)/(2 i)
\end{array}\right) \ \ .
\label{cap5e27}
\eeq
In this parameterization we allow for isospin mixing (the off-diagonal
parameter $a$) and for possible non-unitarity in the $\pi \pi$ two
dimensional subspace (due to opening of other channels).  This is
accomplished by introducing the inelasticity parameters $\eta_{0,2}$
and the extra phase $\Delta$ in the off-diagonal term. The parameters
$\eta_I$ are of order $ 1 + {\cal O} (\alpha^2) $, while $a$ and
$\Delta$ are of order $\alpha$.  

The form given in Eq.~(\ref{cap5e27}) is fully general and includes
all isospin breaking effects. However, strong isospin breaking is
expected to induce only subleading rescattering effects.  In fact,
$m_u - m_d \neq 0$ produces an $I=1$ perturbation to the original
interaction. This is not sufficient to mix the $I=0$ and $I=2$
$\pi\pi$ scattering states when treated to first order, nor does it
cause a splitting of the masses of the charged and neutral pions.
This implies that elastic scattering of these states is still the only
option, and to first order in $m_u - m_d$ the parameter $a$ does not
receive contributions.  
The values of the phases $\delta_I$ may in
principle be slightly modified by the quark mass effect, yet this is
contained in the measured values of the experimental phase shifts.

\subsection{Solution for $\gamma_{0,2}$} 

We are now in position to explore the consequences of unitarity on
the rescattering phases $\gamma_{0,2}$. 
%defined in Eq.~(\ref{cap5e4})
We write 
\beq \overline{\cal A}_{I} =  \overline{A}_{I}  \,
e^{i (\delta_I + \gamma_I )} \ 
\eeq 
and insert these expressions into 
Eq.~(\ref{cap5e25}). We then solve for $\sin \gamma_{0}$ and
$\sin \gamma_{2}$ to first order in $\alpha$,  taking into account 
the $\Delta I = 1/2$ hierarchy of magnitudes.
%
%\beq
%{ {\cal A}_0 \over {\cal A}_2 } R_{ij} \gg R_{ij} \gg
%{ {\cal A}_2 \over {\cal A}_0 } R_{ij} \ \ . 
%\label{app5}
%\eeq
%
After some simple algebra, we obtain the solutions
\beqa
\sin\gamma_0 &=& \beta \left(  {\cal R}e R_{00}  - \tan \delta_0 \, 
{\cal I}m R_{00} \right)  \simeq {\cal O} (\alpha \sin \delta_0) 
\nonumber \\ 
\sin\gamma_2 &=& \beta \, {\overline{A}_0 \over \overline{A}_2 }
\left[ T_{20}  +  \frac{1}{\cos \delta_2} \left( 
 {\cal R}e R_{20} \cos\delta_0  - {\cal I}m R_{20} \sin\delta_0 \right)
 \right]
\label{cap5e28}
\eeqa
The most important feature of these results is the factor
$\overline{A}_0 / \overline{A}_2$ in the formula for $\sin\gamma_2$.
This implies that even though the non-elastic scattering is
electromagnetic in origin, it is enhanced by a large factor that
allows the net change to be significant. Eq.~(\ref{cap5e28}) gives us
the desired expression relating the phase $\gamma_2$ to isospin
breaking rescattering effects.  These are contained in the parameters
$T_{20}$, the mixing amplitude between $\pi \pi$ states, and $R_{20}$.
This last parameter contains the effect of the radiative intermediate
channel $\pi^+ \pi^- \gamma$ as well as the phase space correction.
We note that Eq.~(\ref{cap5e28}) is a generalization of the relation
obtained at one loop in ChPT. However, inspection reveals 
that the perturbative determination contains only the phase space 
effect and the $T_{20}$ mixing in lowest order.

In attempting to estimate the magnitudes of the new phases, we are
hampered by the fact that the analysis of electromagnetic effects in
$\pi\pi$ scattering is not yet complete in the literature. Two groups
have provided analyses of reactions involving neutral 
mesons~\cite{empipi}, but the channels with all charged particles are 
not yet fully analyzed. We require the scattering elements at 
center-of-mass energy equal to the kaon mass. The threshold 
matrix elements are known
from simple tree level calculations, and we will use these in our
estimate below. However, the amplitudes can experience large changes
at $s = M_K^2$, and one needs at least one-loop chiral perturbation
theory in order to obtain these. As these results become available,
they can be used to update our numerical estimates.

We estimate the off diagonal parameter at lowest order in chiral
symmetry obtaining: 
\beq  
T_{02} = { \sqrt{2} \over 3} \cdot {
\delta M_\pi^2 \over 8 \pi F_\pi^2 } \simeq 2.7 \times 10^{-3} \ .
\label{cap5e29}
\eeq
For the parameter $R_{20}$ , proportional to the radiative effect,
one has
\beq R_{20} ={ \sqrt{2} \over 3} \Delta_{+-}
T_{2}^{*} \ ,
\label{cap5e30}
\eeq
and we use the form 
\beq
T_{2} = \frac{1}{\beta} \,  e^{i \delta_2} \,  \sin \delta_2  \ , 
\eeq
with the phenomenological central value of 
$\delta_2 = - 7.0 ^{o} $~\cite{co}.  Numerically this leads to
\beqa
{\cal R}e R_{20}  & = & 0.280 \cdot 10^{-3} \nonumber \\
{\cal I}m R_{20}  & = & 0.034 \cdot 10^{-3}  
\label{cap5e31}
\eeqa
These numerical estimates allow us
to identify the off-diagonal $\Delta I = 2$ rescattering as the major
new ingredient in the final state phases and to arrive at the 
result: 
\beq 
\gamma_0 = - 0.1 ^{o}    \ , \qquad 
\gamma_2  =  3.1^{o} \ \ .
\label{cap5e32}
\eeq
%\gamma_0 = {\cal O}( \alpha) \simeq 0 \ , \qquad \gamma_2
We note here that the result for $\gamma_2$ is quite large, 
amounting to almost $50 \%$ of the strong phase $\delta_2$ at 
$s = M_{K}^{2}$.

\section{Sample Fit to  $K \rightarrow \pi \pi$ Data}
\label{fit}

In this section, we provide a tentative fit to the present experimental 
data. This is meant as an illustration of the ideas that we have discussed
above, and hopefully will provide a model for a new fully consistent 
experimental analysis of new data, taken with the full treatment of 
electromagnetic effects. We describe our treatment as tentative because
it involves older data sets which were taken without the inclusion of
radiative corrections. We cannot fully account for the experimental
acceptances, and are forced to adopt a cruder procedure. However, the
sample fit is none the less of interest because it illustrates the 
significant sensitivity
of various quantities to electromagnetic corrections, and represents the
best that can be done with the present data set.

\subsection{Data Analysis}

It will be convenient in the discussion to follow to first define
\beq
\chi_i \ \equiv \ \delta_i + \gamma_i \qquad (i = 0,2) \ \ .
\label{chi}
\eeq
Then having $G_{+-} (\omega)$, $G_{+0} (\omega)$ and the structure
dependent corrections $\delta A_{I}^{\rm EM}$, one is in a position 
by using Eqs.~(\ref{cap5e2}) and (\ref{cap5e4})
to extract the quantities $A_0$, $A_2$ and $\chi_0 - \chi_2$.  

\begin{figure}[htb]
\centering
\begin{picture}(200,150)  
\put(50,20){\makebox(100,120){\epsfig{figure=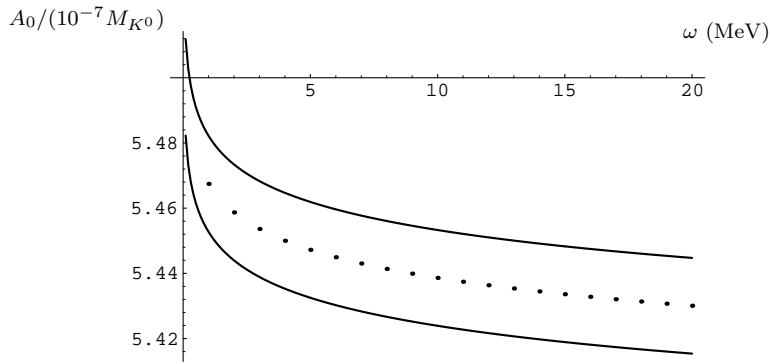,height=3in}}}
\put(200,140){\scriptsize{$\omega$ (MeV)}}
\put(-55,145){\scriptsize{$A_0 / (10^{-7} M_{K^0})$}} 
\end{picture}
\caption{Fitted $A_0$ as a function of $\omega$ .\hfill 
\label{cap5fig3}}
\end{figure}

\begin{figure}[htb]
\centering
\begin{picture}(200,150)  
\put(50,20){\makebox(100,120){\epsfig{figure=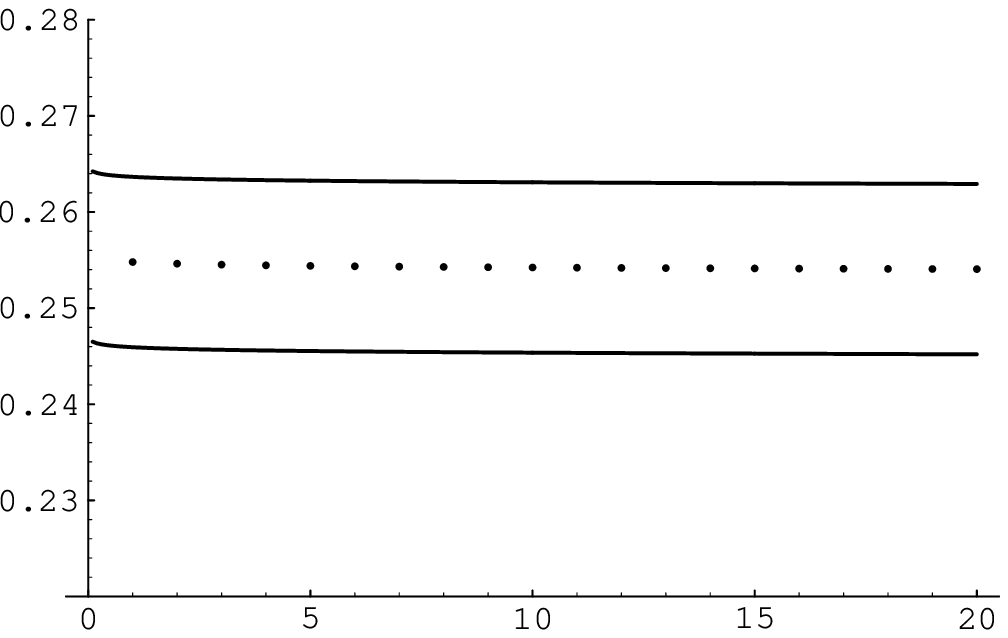,height=3in}}}
\put(200,0){\scriptsize{$\omega$ (MeV)}}
\put(-75,145){\scriptsize{$A_2 / (10^{-7} M_{K^0})$}} 
\end{picture}
\caption{Fitted $A_2$ as a function of $\omega$ .\hfill 
\label{cap5fig4}}
\end{figure}
\begin{figure}[htb]
\centering
\begin{picture}(200,150)  
\put(50,20){\makebox(100,120){\epsfig{figure=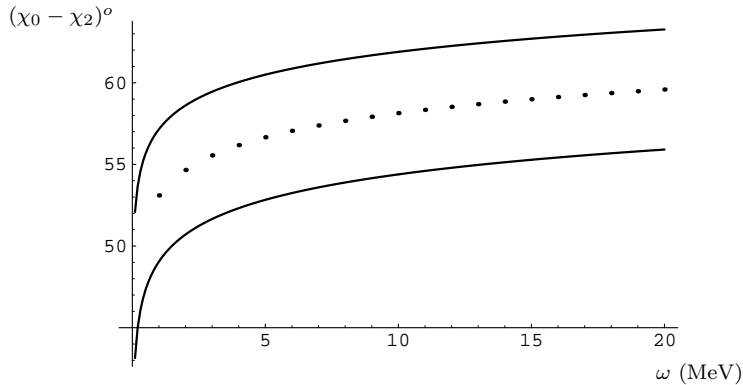,height=3in}}}
\put(200,10){\scriptsize{$\omega$ (MeV)}}
\put(-45,145){\scriptsize{$(\chi_0 - \chi_2)^{o}$}} 
\end{picture}
\caption{Fitted $\chi_0 - \chi_2$ as a function of $\omega$ .\hfill 
\label{cap5fig5}}
\end{figure}
As experimental input for the branching ratios we use the PDG
averages, although these numbers come with no reference to what
portion of the $K \rightarrow \pi \pi
\gamma$ mode is included.  In order to understand the attendant 
uncertainties and ambiguities of this approach, in Figs.~\ref{cap5fig3},
\ref{cap5fig4} and \ref{cap5fig5} we plot the output of our fit as a
function of the parameter $\omega$, the upper cutoff for IR photons in
the center of mass frame. 
It is not clear to which value of $\omega$ (if any) the experimental
numbers correspond. This ignorance gives rise to little  
uncertainty in the extraction $A_2$ and to a moderate one in the 
extraction of $A_0$, for $\omega$ varying
between 1 MeV and 20 MeV.\footnote{This range is chosen to 
reflect a realistic possibility for detector resolution.}
 However, more delicate is the situation for
the extraction of the phase $\chi_0 - \chi_2$, where a variation 
of the order of $10 \%$ is seen over the considered range of $\omega$. 
Thus our analysis indicates that the extraction of rescattering phases 
from $K \rightarrow \pi \pi$ data is sensitive to the treatment of
soft photons. In the absence of precise experimental information, 
it is not possible to pick a definite central value for our output.   
We thus quote the results for the set of EM-free quantities with two error 
bars.  The first one is due to the spread in the central values
according to variations of $\omega$ between 1 and 20 MeV. The second
one comes from propagating the experimental uncertainty in the decay 
widths and the theoretical uncertainty on the inputs $\delta A_{I}^{\rm EM}$. 
%As a consequence, the error bar on
%the quantity $\chi_0 - \chi_2$ has to be considerably increased. 
We find:
\begin{eqnarray}
 A_{0} & = & (5.450   \pm   0.020 \pm 0.015) \times 10^{-7} M_{K^{0}} \ , 
\nonumber \\ 
 A_{2} & = & (0.255  \pm   0.001 \pm 0.009) \times 10^{-7} M_{K^{0}} \ , 
\label{fitaverage}  \\
 \chi_{0} - \chi_{2} & = & (56  \pm 4 \pm 4)^{o} \ .  \nonumber
\end{eqnarray}
These results should be compared with the ones presented in
Eq.~(\ref{cap0e3}), derived from the analysis in the isospin limit.
The most important new feature is that considering EM corrections places
larger error bars on all these quantities.  In the case of $A_2$ the
reason for this resides in the quite large theoretical uncertainty on
$\delta A_{2}^{+}$.  For $A_0$ and the phase difference $\chi_0 -
\chi_2$, the larger error bar is due essentially to incomplete
information concerning the treatment of the radiative channel. A
measurement of the the partial width $\Gamma_{+-} (\omega)$, with
accuracy level of $\sim 0.5 \%$ (this is the accuracy level
of the present PDG numbers), accompanied by information on
soft-photon cuts, would allow one to extract a definite central value
for $A_0$ and $\chi_0 - \chi_2$.  As a consequence, this would
eliminate the first error bar associated with $A_0$ and $\chi_0 -
\chi_2$ in Eq.~(\ref{fitaverage}), reducing the total uncertainty 
by $50 \%$ or more. Indeed, such an analysis will be performed by the KLOE 
experiment at Da$\Phi$ne \cite{franzini}.  

\subsection{Extraction of $\delta_0 - \delta_2$: Discussion}

Finally we turn to the extraction of  $\delta_0 - \delta_2$ 
from $K \rightarrow \pi \pi$ data. The relation to be used is:
\beq
\delta_0 - \delta_2 \  =  \ \left( \chi_0 - \chi_2 \right)_{\rm fit} \ + \ 
	\gamma_2 - \gamma_0  \ .
\label{cap5e33}
\eeq
As shown by Eq.~(\ref{cap5e33}), the extraction of the strong phase 
difference relies on two distinct inputs:
\begin{enumerate}
\item The first one comes from the fit to $K \rightarrow \pi \pi$ 
branching ratios, which provides $\chi_0 - \chi_2$.  In Sect.~\ref{fit}A 
we discussed such a fit and pointed out the sensitivity 
of $\chi_0 - \chi_2$ to cuts used for soft real photons. 
In the absence of information on these cuts a precise determination of 
$\chi_0 - \chi_2$ is not possible and our conclusion is that 
the error bars are larger than previously thought: 
\beq
\chi_0 - \chi_2 = \left(56 \pm 8 \right)^{o} 
\label{cap5e34}
\eeq
\item The second input concerns the magnitude of $\gamma_2 -
\gamma_0$. We have thus far established a general framework 
(based on unitarity) for the analysis of 
these isospin breaking phases. We 
found that $\gamma_2$ receives a $\Delta I = 1/2$ enhancement 
and the dominant effect is due to isospin mixing in the $\pi
\pi$ rescattering rather than to radiative intermediate channels. 
We have provided an estimate of $T_{20}$ at lowest order in 
the chiral expansion, leading us to write:
\beq
\gamma_2 - \gamma_0 = 3.2^{o} + \gamma_{2}^{(e^2 p^2)} \ . 
\label{cap5e35}
\eeq
The possibility of large chiral corrections to $T_{20}$ 
(associated with 
$\gamma_{2}^{(e^2 p^2)} $) cannot be ruled out, given the results obtained 
in the analysis of other EM corrections (violations of 
 Dashen's theorem and the $K  \rightarrow \pi \pi$ amplitudes). 
\end{enumerate}
In light of the previous discussion, we give the following value for 
$\delta_0 - \delta_2$ from $K \rightarrow \pi \pi$ data:
\beq
\delta_0 - \delta_2 =  \left( 59  + \gamma_{2}^{(e^2 p^2)} \pm 8 
\right)^{o} \ .  \label{cap5e36}
\eeq
The leading order estimate for $\gamma_2$ is seen to worsen the
discrepancy between the central values of weak and strong
determinations of $\delta_0 - \delta_2$. However, the large
uncertainty associated with radiative corrections makes impossible a
precise comparison at this stage.  In this sense, the phase puzzle is
alleviated, its cause being a previous underestimate of error bars.
Indeed we believe that the combined effect of radiative corrections to
$\chi_0 - \chi_2$ and calculation of $\gamma_2^{(e^2 p^2)}$ can fully
resolve the puzzle, providing a satisfactory theoretical formulation
of the problem.  In fact, once a more precise extraction of $\chi_0 -
\chi_2$ becomes available, Eq.~(\ref{cap5e33}) can be used to extract
$T_{20}$, and thus information on the isospin breaking dynamics in
$\pi \pi$ scattering at $s = M_{K}^{2}$.

\section{Impact on CP Phenomenology}
\label{CPV}

In the present section we focus on the consequences of our work to CP
phenomenology in the kaon system.  Our work gives rise to interesting
effects only in the theoretical analysis of $\epsilon '$. In
particular, we provide an estimate of the isospin breaking
parameter $\Omega^{\rm EM}$, the effect of the $\Delta I = 5/2$ amplitude
and the phase of $\epsilon '$.

The analysis of direct CP-violation in $K \rightarrow \pi \pi$ 
 proceeds exactly as in the standard case,
except that now we work with the IR finite isospin amplitudes
$\overline{A}_{I}$ and the final state interaction phases $\chi_I$
associated with them.
One can then write 
\beq
\epsilon '  = - \frac{i}{\sqrt{2}} \, e^{i ( \chi_2 - \chi_0 )} \, 
\frac{{\cal R}e \overline{A}_{2}}{ {\cal R}e \overline{A}_{0}} \,  
\left[ 
\frac{{\cal I}m \overline{A}_{0}}{ {\cal R}e \overline{A}_{0}} \, - \,  
\frac{{\cal I}m \overline{A}_{2}}{ {\cal R}e \overline{A}_{2}} \right]  
\ . 
\label{cp8}
\eeq
Defining 
\beq
\overline{\omega} = 
\frac{{\cal R}e \overline{A}_{2}}{{\cal R}e \overline{A}_{0}} \ , 
\eeq
and neglecting the small effect of $\delta A_0 / A_0$ one arrives at 
\beq
\epsilon '  = - \frac{i}{\sqrt{2}} \, e^{i ( \chi_2 - \chi_0 )} \, 
\overline{\omega} \, 
\frac{{\cal I}m A_{0}}{{\cal R}e A_{0}} \, 
\left[ 1 \,  - \, \frac{1}{\overline{\omega}} \,  
\frac{{\cal I}m \overline{A}_{2}}{ {\cal I}m {A}_{0}}  \right]  \ . 
\label{cp8.5}
\eeq
We recall here that in the Standard Model analysis  
the imaginary part of $A_0$ is generated by the so
called gluonic penguin, while the phase of $A_2$ is
generated by the electroweak penguin. 

In order to make manifest the effects of electromagnetic corrections,
we now further study Eq.~(\ref{cp8.5}). 
The first new effect is to be found in the parameter
$\overline{\omega}$.  It is due to the presence of the $\Delta I =
5/2$ amplitude, distinguishing $\overline{A}_{2}$ from
$\overline{A}_{2}^{+}$ (see Eq.~(\ref{cap5e4})).  In the usual
treatment one uses the parameter 
\beq
\omega = 
\frac{{\cal R}e \overline{A}^{+}_{2}}{{\cal R}e \overline{A}_{0}} 
 = \frac{1}{22.2}   \ . 
\eeq   
However, our derivation shows that one should use $\overline{\omega}$. 
The two are related by:
\beq
\overline{\omega} = 
\frac{{\cal R}e \overline{A}^{+}_{2}}{{\cal R}e \overline{A}_{0}} 
\frac{{\cal R}e \overline{A}_{2}}{{\cal R}e \overline{A}^{+}_{2}}  = 
 \omega \left( 1 + f_{5/2} \right)  \ .  
\label{cp8.75}
\eeq   
%
%
%\beq
%{\cal R}e \overline{A}_{2} = {\cal R}e \overline{A}^{+}_{2} \, 
%\left( 1 + f_{5/2} \right) 
%\label{cp9}
%\eeq

The other relevant phenomenon is the leakage of the octet amplitude
into $\overline{A}_{2}$, providing the dominant part of $\delta
{A}_{2}$.  This brings an extra contribution to the CP-violating phase
of $\overline{A}_{2}$, essentially generated by the gluonic penguin
and transferred to $\overline{A}_2$ via isospin breaking effects. This
mechanism is usually parameterized by:
\beq
\Omega^{\rm iso-brk} = 
\frac{1}{\omega}  
\, \frac{ {\cal I} m \, \delta {A}_{2}^{\rm iso-brk} }{ 
{\cal I} m {A}_{0}}  \ \ , 
\label{cp10}
\eeq
where $\Omega^{\rm iso-brk}$ will have contributions from both 
electromagnetic effects ($\Omega^{\rm EM}$) and from 
strong interaction effects ($\Omega^{\rm STR}$)
associated with $m_u \ne m_d$,  
\beq
\Omega^{\rm iso-brk} \equiv \Omega^{\rm EM} + 
\Omega^{\rm STR} \ \ .
\label{omega}
\eeq
The above observations lead us to write:
\beq
\epsilon '  = - \frac{i}{\sqrt{2}} \, e^{i ( \chi_2 - \chi_0 )} \, 
\omega \, \frac{{\cal I}m {A}_{0}}{ {\cal R}e {A}_{0}} \, 
\left[ 1 - \frac{1}{\omega} \, \frac{{\cal I}m {A}_{2}}{ {\cal I}m {A}_{0}} + 
f_{5/2} - \Omega^{\rm iso-brk}   \right] 
\label{cp11}
\eeq
Comparing Eq.~(\ref{cp11}) to the standard analysis (not including EM 
corrections), one identifies three new effects. 

\begin{enumerate}

\item The factor $f_{5/2}$ appears: it  can be obtained by
inserting in Eq.~(\ref{cp8.75}) our previous estimates of $\delta
A_{2}$ and $\delta A_{2}^{+}$ (see Ref.~\cite{em3}). 
We find 
\beq
f^{\rm EM}_{5/2}  =  \left( 9.3  \pm 6.1 \right) \cdot 10^{-2}   \  . 
\label{cp12}
\eeq
The large uncertainty reflects the one in $\delta A_{2}^{+}$. 
We thus find that this effect tends to increase (although slightly) 
the central value of $\epsilon' / \epsilon$. 
The authors of Ref.~\cite{valencia} find an opposite result because they 
use the ``phenomenological'' value of $A_{5/2}$. 
We believe that the phenomenological determination of $A_{5/2}$, 
as performed in Ref.~\cite{valencia},  suffers from large systematic 
uncertainties due to neglecting IR effects and the EM phases $\gamma_I$. 

\item One has to consider the electromagnetic contribution 
$\Omega^{\rm EM}$, to be added to existing estimates of 
$\Omega^{\rm STR}$ due to strong isospin breaking. Again, 
the analysis performed in Ref.~\cite{em3} 
enables us to get the magnitude of $\Omega^{\rm EM}$, since 
we calculated there the octet induced component of 
$\delta A_{2}^{\rm EM}$. Thus we can write:
\beq
\Omega^{\rm EM} =  
 \frac{ {\cal R}e \, {A}_{0}}{ {\cal R} e {A}_{2}}  \cdot 
\frac{ {\cal I}m \,  \delta {A}_{2}}{ {\cal I}m  {A}_{0}}  = 
\frac{ {\cal R}e \, {A}_{0}}{ {\cal R} e {A}_{2}}  \cdot 
\frac{ {\cal R}e \, \delta {A}_{2}}{ {\cal R} e {A}_{0}}   \ . 
\label{cp13}
\eeq
Numerically we find: 
\beq
\Omega^{\rm EM} = \left( 6.0  \pm 2.5 \right) \cdot 10^{-2} \ .   
\label{cp14}
\eeq
\item One observes that the phase of $\epsilon '/ \epsilon$ is related
to $\chi_0 - \chi_2$ and not to $\delta_0 - \delta_2$, although with
the present accuracy it is hard to make a meaningful determination. We
find
\beq
\Phi^{\epsilon' /\epsilon} = \left( \chi_2 - \chi_0 + \frac{\pi}{2} 
\right) -  \frac{\pi}{4}   = - \left( 11 \pm 8 \right)^{o} \ .   
\eeq
The resulting effect on the real and imaginary part of $\epsilon ' / 
\epsilon$ is below the sensitivity of present  kaon factories. 

\end{enumerate}

We conclude by observing that the individual terms $f_{5/2}$ and 
$\Omega^{\rm EM}$ have a respectable size but enter in the expression 
for $\epsilon '$ with opposite sign. The net effect has a very small 
central value with a large uncertainty.

\section{Conclusions} 

In this paper we have attempted a full phenomenological analysis of $K
\rightarrow \pi \pi$ decays in the presence of electromagnetic
interactions.  We have provided a general parameterization of $K
\rightarrow \pi \pi$ amplitudes to include the effect of isospin breaking
interactions. Such a parameterization has allowed us to organize the
calculation in terms of three main effects: structure dependent
corrections (see Refs.~\cite{em1,em2,em3}), electromagnetic 
infrared corrections, and isospin breaking in final state 
interactions.  We have also studied the effect of electromagnetic
corrections on the direct CP-violation parameter $\epsilon '$.

\subsection{IR Effects: Need for New B.R. Measurements} 

It is well known that the calculation of IR effects requires knowledge 
of the experimental cuts used in treating the soft photons emitted in  
the $K \rightarrow \pi \pi$ decays. In Sect.~\ref{fit} we have pointed 
out that the PDG numbers come with no information concerning the 
radiative channel, and this seriously compromises any attempt to 
properly include the radiative corrections.  In the absence of 
experimental input, we have performed a calculation of the IR effects in a
simple theoretical scheme (isotropic cut on the photon energy in the
center of mass system).  We have shown how this incomplete
state of affairs produces uncertainties larger than previously 
thought in the EM-free quantities

We strongly urge that a measurement of the $K \rightarrow \pi
\pi$ branching ratios be performed at one of the current 
high statistics kaon
experiments. To be precise, it would be interesting to have a set of
measurements of $\Gamma_{n} (\omega)$ ($n = +-, +0$) at different
values of $\omega$ (the soft photon upper cutoff in the center of mass
frame). This would allow anyone to apply our calculation of
$G_{+-}$ and $G_{+0}$ in making a phenomenological analysis (as
in Sect.~\ref{fit}).  Of course, each distinct experimental procedure would
require its own theoretical calculation of $G_{+-,+0}$. All such
studies would be equally welcome, as long as they 
provide information on the inclusive
sum of $\pi \pi$ and $\pi \pi \gamma$ channels.  We stress that such 
measurements are necessary in order to fully address the impact of EM on
$K \rightarrow \pi \pi$ decays.

\subsection{Final State Interaction Phases}  

We have shown that isospin breaking changes the description of
rescattering phases in $K \rightarrow \pi \pi$ decays, as Watson's 
theorem is no longer applicable. We have described this new feature
within the general framework provided by the unitarity relations,
pointing out that the relevant effect is of electromagnetic origin. In
Sect.~\ref{phases} we have set up the framework relating the extra
phases $\gamma_{0,2}$ to EM effects in $\pi \pi$ scattering. Our
leading order analysis finds a large effect 
in $\gamma_2$, equal to $50 \%$ of
the strong phase $\delta_2$. The general framework presented has the
potential to fully resolve the long standing inconsistency between the
strong determination of $\delta_0 - \delta_2$ at $s = M_K^2$ and the
one emerging from $K \rightarrow \pi \pi$ data. At present, little can
be concluded due to the large uncertainty in the phase $\chi_0 -
\chi_2$ and the lack of a calculation for $\gamma_{2}^{(e^2 p^2)}$.
The first problem will be solved by new measurements of the branching
ratios (including proper information on radiative effects).  The 
second problem depends on the theoretical ability to calculate EM 
corrections to $\pi \pi$ scattering at order $e^2 p^2$ in the chiral 
expansion.  

\subsection{CP Phenomenology}  

Finally, we have analyzed the impact of electromagnetic corrections on
CP phenomenology (see Sect.~\ref{CPV}), pointing out the new 
features in the study of  $\epsilon ' / \epsilon$. 
The isospin breaking effects can be encoded into the factors
$\Omega$ and $f_{5/2}$, and also affect the phase of $\epsilon '$.
Both $f_{5/2}$ and $\Omega$ receive contributions from strong isospin
breaking and electromagnetism.  We have provided an estimate for the
electromagnetic effect, finding results of the order of $10 \%$, for
these parameters. They appear with opposite sign, and thus do not
produce sizeable shifts in the theoretical prediction of $\epsilon ' /
\epsilon$.

\acknowledgments 

This work was supported in part by the National
Science Foundation.  One of us (V.C.) acknowledges 
support from the Foundation A. Della Riccia.

\eject

\eject


\begin{thebibliography}{99}

\bibitem{em1} V. Cirigliano, J.F. Donoghue
and E. Golowich, Phys. Lett. {\bf B450}  (1999) 241.

\bibitem{em2} 
%{\it Electromagnetic Corrections to $K \to \pi\pi$ I --
%Chiral Perturbation Theory}, 
V. Cirigliano, J.F. Donoghue and E. Golowich, Phys.Rev. {\bf D 61}:
093001, 2000.

\bibitem{em3} 
%{\it Electromagnetic Corrections
%to $K \to \pi\pi$ II -- Dispersive Matching},
V. Cirigliano, J.F. Donoghue and E. Golowich:  Phys.Rev. {\bf D 61}:
093002, 2000. 

\bibitem{PDG} PDG98, C. Caso {\it et al.}, Eur.Phys.J. {\bf C3} (1998) 1.


\bibitem{old}   F. Abbud, B.W. Lee and C.N. Yang,
Phys.Rev.Lett.{\bf {18}} (1967) 980 ; \\ A.A. Belavin and I.M. Narodetskii,
Sov.J.Nucl.Phys.{\bf 8} (1968) 568; \\ A. Neveu and J. Scherk, Phys.Lett.{\bf 
B27} (1968) 384; \\  A.A. Bel'kov and V.V. Kostyuhkin,
Sov.J.Nucl.Phys. {\bf 51} (1989) 326. 

\bibitem{der} E.de Rafael, Nucl.Phys. {\bf B 7A} (Proc. Suppl.) (1989) 1. 

\bibitem{Vienna00} G. Ecker, G. Isidori, H. Neufeld, G. Muller, 
A. Pich, 
%{\em Electromagnetism in Nonleptonic Weak Interactions}, 
hep-ph/0006172. 

\bibitem{isob}  J.F. Donoghue, E. Golowich, B.R. Holstein, J. Trampetic, 
Phys.Lett.{\bf B179} (1986) 361; \\
A.J. Buras, J.M. Gerard, Phys.Lett.{\bf B192} (1987) 156; \\ 
H.Y. Cheng, Phys.Lett.{\bf B201} (1988) 155.

\bibitem{isobnew} 
S. Gardner, G. Valencia, Phys.Lett.{\bf B466} (1999) 355; \\ 
G. Ecker, G. Muller, H. Neufeld, A. Pich, Phys.Lett.{\bf B477} (2000) 88; \\ 
K. Maltman, C. Wolfe,  Phys.Lett.{\bf B482} (2000) 77. 
%hep-ph/9912254.

\bibitem{valencia} S. Gardner, G. Valencia,  hep-ph/0006240.

%\bibitem{devdi} T.J.Devlin, J.O. Dickey, Rev.Mod.Phys. {\bf 51} (1979) 237. 

\bibitem {acgl} B. Ananthanarayan, G. Colangelo, J. Gasser and
H. Leutwyler, hep-ph/0005297.

\bibitem{GassMeiss} J. Gasser and U. Meissner, Phys.Lett.{\bf B258} (1991) 219.

\bibitem{F61} D.R. Yennie, S.C. Frautschi, H. Suura, Ann.Phys. (NY) {\bf 13}
 (1961) 379.

\bibitem{Weinberg65} S. Weinberg, Phys.Rev. {\bf 140} (1965) 516.

\bibitem{franzini} Talk given by P.Franzini at {\em Chiral 2000}, 
JLAB, July 17-22 2000. 

\bibitem{empipi} U. Meissner, G. Muller, S. Steininger, Phys.Lett.{\bf B406} 
(1997) 154, Erratum-ibid.{\bf B407} (1997) 454 ; \\
M. Knecht, R. Urech, Nucl.Phys.{\bf B519} (1998) 329. 



\end{thebibliography}
\end{document}